\documentstyle[aps]{revtex}
\begin{document}
\draft

\topmargin.2cm
\preprint{\vbox{\hbox{CINVESTAV-FIS-15/96}}}

\title{Lepton mass generation and family number violation mechanism in the
       $\bbox{\rm SU(6)_L \otimes U(1)_Y}$ model}
\author{Umberto Cotti}
\address{SISSA-ISAS, via Beirut 2-4, 34010 Trieste, Italia%
         \thanks{and Departamento de F\'{\i}sica, Centro de 
         Investigaci\'on y de Estudios Avanzados del IPN A.P. 14-740, 
         07000 M\'exico D.F., M\'exico.}
         }
\author{Ricardo Gait\'an\thanks{e-mail: rgaitan@servidor.unam.mx}}
\address{Centro de Investigaciones Te\'oricas, Facultad de Estudios
         Superiores-Cuautitl\'an,
         UNAM, A.P. 142, 54700 Cuatitl\'an Izcalli,
         Estado de M\'exico, M\'exico.}
\author{A. Hern\'andez-Galeana}
\address{Departamento de F\'{\i}sica, Escuela Superior de F\'{\i}sica y
        Matem\'aticas del Instituto Polit\'ecnico 
        Nacional, 07738 M\'exico D.F., M\'exico.}
\author{William A. Ponce}
\address{Departamento de F\'{\i}sica, Universidad de Antioquia 
         A.A. 1226, Medell\'{\i}n, Colombia}
\author{Arnulfo Zepeda} 
\address{Departamento de F\'{\i}sica, Centro de Investigaci\'on 
         y de Estudios Avanzados del IPN A.P. 14-740, 
         07000 M\'exico D.F., M\'exico.}
\date{Abril 21, 1998}
\maketitle
\begin{abstract}
 Lepton family number violation processes arise in the 
$\rm SU(6)_L \otimes U(1)_Y$ model due to the presence of an extra neutral
gauge boson, Z$'$, with family changing couplings, and due to
the fact that this model demands the existence of heavy exotic leptons. 
The mixing of the standard Z with Z$'$ and the mixing of ordinary leptons
with  exotic ones induce together family changing couplings on the Z and 
therefore nonvanishing rates for lepton family number violation processes,
such
as 
$\rm Z \rightarrow e \bar{\mu}$, 
$\rm \mu \rightarrow ee\bar{e}$ and 
$\rm \mu \rightarrow e\gamma$. Additional contributions to the processes
$\rm \mu \rightarrow e \gamma$ and $\rm \mu \rightarrow ee \bar{e}$ are
induced from the mass generation mechanism. This last type of
contributions
may compete with the above one, depending on the masses of the scalars
which participate in the diagrams which generate radiatively the masses of
the charged leptons. Using the experimental data we compute some bounds
for the mixings parameters and for the masses of the scalars.   
\end{abstract}


\section{Introduction}
\begin{sloppypar}
 In this communication we show how the tree level 
{\sl family changing neutral current} (FCNC) phenomenon arises 
in the context of the $\rm SU(6)_L \!\otimes\! U(1)_Y$ 
model ~\cite{zpc55:423} whose simplicity allows to tests several 
mechanisms for physics beyond the Standard Model (SM).  
 Within this model we then compute the branching ratios for
{\sl lepton family number violating} (LFV) processes in the e--$\mu$
sector  
with the aim of bounding the light-heavy lepton mixing parameters of 
the model in this sector.
 In section~\ref{model} we present the main features of the 
$\rm SU(6)_L \!\otimes\! U(1)_Y$  model and in section~\ref{symm-break}
those of its symmetry breaking.
 In section~\ref{Zomix} and~\ref{mixord} we display the mixing effects 
in the leptonic sector and the FC couplings of the Z to ordinary 
leptons; in section II E we deal with the mechanism which gives masses
radiatively to the charged leptons and in section~\ref{constr} we compute
the branching ratio for 
$\rm Z \rightarrow e \bar{\mu}$ and $\rm \mu \rightarrow ee\bar{e}$, both
at 
tree level and for $\rm \mu \rightarrow e\gamma$ at one loop. 
\end{sloppypar}

\section{The model}
\subsection{Group Structure}
\label{model}
 The gauge group of the model is $\rm SU(6)_L \!\otimes\! U(1)_Y$, where 
$\rm SU(6)_L$ unifies the weak
isospin $\rm SU(2)_L$, of Glashow-Weinberg-Salam, with an horizontal 
gauge group 
$\rm G_H$, whose maximum expression is $\rm SU(3)_H$. $\rm SU(2)_L
\!\otimes\! SU(3)_H$
is a maximal special 
subgroup of $\rm SU(6)_L$ .
The thirty-five $\rm SU(6)_L$ generators, {\sf S}$_a$,  in the 
$\rm SU(2)_L \!\otimes\! SU(3)_H$ basis are
\begin{equation}
{\sf S}_a  =  \hspace{8mm} 
 \case{1}{2\sqrt{3}} {\sigma}_i \otimes {\sf 1}_3 
  \longleftrightarrow {\rm W}_i, \hspace{8mm}
 \case{1}{2\sqrt{2}} {\sf 1}_2 \otimes {\lambda}_\alpha  
  \longleftrightarrow {\rm G}_\alpha, \hspace{8mm}
 \case{1}{4}{\sigma}_i \otimes {\lambda}_\alpha 
  \longleftrightarrow  {\rm H}_{i\alpha},
\end{equation}
where ${\sigma}_i$, $i$ = 1,2,3, are the Pauli matrices, 
${\sf T}_i$ = $\case{1}{2}{\sigma}_i$ are the $\rm SU(2)_L$ generators, 
${\lambda}_\alpha$, $\alpha=1,2,\ldots,8$, are the Gell-Mann
matrices, 
$\case{1}{2}{\lambda}_\alpha$ are the $\rm SU(3)_H$ generators, and 
{\sf 1}$_3$
and {\sf 1}$_2$ are 3$\times$3 and 2$\times$2 unit matrices, respectively. 
The corresponding gauge bosons are indicated. Notice that all the 
generators are equally normalized:
\begin{equation}
 {\rm Tr} {\sf S}_a{\sf S}_b = \case{1}{2}\delta_{ab}, \hspace{4mm} 
 {\rm Tr} {\sf T}_i {\sf T}_j = \case{1}{2}\delta_{ij}, \hspace{4mm} 
 {\rm Tr} \case{1}{4}{\lambda}_{\alpha} {\lambda}_{\beta} = 
  \case{1}{2}\delta_{\alpha\beta} 
\end{equation}
with the consequence that if $g_6$, $g_2$ and $g_{\rm H3}$ are the coupling 
constants of $\rm SU(6)_L$, $\rm SU(2)_L$ and $\rm SU(3)_H$, respectively, 
then they satisfy, at the unification scale, the relationships
\begin{equation}
 g_2 = \case{1}{\sqrt{3}}g_6, \hspace{3cm} g_{\rm 3H} = \case{1}{\sqrt{2}} g_6.
\label{g2g3H}
\end{equation}
The $g$ in the standard notation of the Standard Model, is $g = g_2$.

The model has 36 gauge bosons: 
35 associated with the generators of $\rm SU(6)_L$ and one associated to 
$\rm U(1)_Y$. 
Besides the standard gauge bosons  $\rm W^3$, $\rm W^{\pm}$ and B, we 
have 32 extra
gauge bosons, which can be divided into four groups: 
\begin{quote}
(++), {\sf 12 charged gauge bosons} which perform transitions among 
 families. They couple to  {\sl family changing charged currents} (FCCC)
and they are related to ${\sigma}_i \otimes {\lambda}_\alpha$, 
$i=1,2$, $\alpha = 1,2,4,5,6,7.$

(+\,o), {\sf 4 charged gauge bosons} which do not make transitions among 
 families but their couplings are family dependent. They couple to 
 {\sl non-universal family diagonal charged currents} (NUFDCC) and are 
 related to  
${\sigma}_i \otimes {\lambda}_\alpha$, $i=1,2$,
$\alpha = 3,8.$

(o\,+), {\sf 12 neutral gauge bosons} 
 which induce transitions among families. They couple to FCNC and are 
 related to ${\sf 1}_2\otimes{\lambda}_\alpha$,
${\sigma}_3 \otimes {\lambda}_\alpha$,  $\alpha = 1,2,4,5,6,7.$

(o\,o), {\sf 4 neutral gauge bosons} which couple non-universally without 
 changing flavor, that is they couple to {\sl non-universal family 
diagonal neutral currents} (NUFDNC) and are related to 
${\sf 1}_2\otimes{\lambda}_\alpha$,
${\sigma}_3\otimes{\lambda}_\alpha$,  $\alpha = 3,8$. 
\end{quote}
 Notice that at scales where families are not defined, that is at scales 
above the breaking of $\rm SU(2)_L\otimes U(1)_Y$, there is no clear 
distinction between the members of the (+\,o) set and those of the (++) one. 
 Similarly, the members of the set (o\,+) and those of the set (o\,o) 
should be put in a unique class. 
 In other words, there is no way to distinguish ``family changing" from 
``non-universal flavor diagonal" attributes. 
 For example a gauge boson associated with a generator of the form
\begin{equation}
{\sf T}_{\rm H1} = \frac{1}{\sqrt2}\left(\begin{array}{ccc}
                0 & 1 & 0 \\
                1 & 0 & 1 \\
                0 & 1 & 0 
                \end{array}  \right),
 \label{HO}
\end{equation}
in family space, will be identified as a family changing one, but applying 
a transformation in family space one can diagonalize {\sf T}$_{\rm H1}$ to
the form
\begin{equation}
{\sf T}^{\prime}_{\rm H1} = {\sf T}_{\rm H3} = \left(\begin{array}{rrr}
            1 & 0 & 0\\
            0 & 0 & 0\\
            0 & 0 & -1
            \end{array}  \right)
\end{equation}
and therefore transform the corresponding gauge boson into one with 
non-universal flavor diagonal couplings.
 
The fermionic content of the model is given by the following set of
irreducible representations (irreps) of $\rm SU(6)_L\!\otimes\!U(1)_Y$, 
one for each color in the case of quarks (right handed fields are
charge-conjugated to left handed ones):
\begin{tabbing}
 $\{{\bf 1}_I(-\frac{4}{3})\}_{\rm L}$ \= =  
   $\rm (e^-,\nu_e,\mu^-,\nu_{\mu},\tau^-,\nu_{\tau})_L$\ \ \              
  \= $\equiv\;\;\psi_{[\alpha\beta]}(0)_L$,               \=  \kill  \\
 $\{{\bf 6}(\frac{1}{3})\}_{\rm L}$    \> =~ $\rm (u,d,c,s,t,b)_L$\                   
  \> $\equiv\;\;\psi^{\alpha}_{\rm L}(\frac{1}{3})$,        \>         \\[1mm]
 $\{\overline{\bf 6}(-1)\}_{\rm L}$         \> =~ 
   $\rm (e^-,\nu_e,\mu^-,\nu_{\mu},\tau^-,\nu_{\tau})_L$\                  
  \> $\equiv\;\;\psi_{\alpha {\rm L}}(-1)$,                 \>         \\[1mm]
 $\{{\bf 1}_I(-\case{4}{3})\}_{\rm L}$ \> =~ 
  $q^{\rm c}_{I \rm L}(-\case{4}{3})$ 
  \> \> \` $I$~=~1,2,3 for $\rm u^c$, 
   $\rm c^c$ and $\rm t^c$ respectively,      \\[1mm]
 $\{{\bf 1}_I(\frac{2}{3})\}_{\rm L}$  \> =~ 
   $q^{\rm c}_{I \rm L}(\case{2}{3})$
  \> \> \` $I$~=~1,2,3 for $\rm d^c$, $\rm s^c$ and 
   $\rm b^c$ respectively,      \\[1mm]
 $\{{\bf 1}_I(2)\}_{\rm L}$            \> =~ $l^{\rm c}_{I \rm L}(2)$            
  \> \> \` $I$~=~1,2,3 for $\rm e^c$, 
   $\rm \mu^c$ and $\rm \tau^c$ respectively, \\[1mm]
 $\{\overline{\bf 15}(0)\}_{\rm L}$         \>                               
  \> $\equiv\;\;\psi_{[\alpha\beta] {\rm L}}(0)$,           \>
\end{tabbing}
\noindent where $I$ is a family index, $\alpha$ and $\beta$ are 
$\rm SU(6)_L$ tensor
indices and u, d, \ldots\ refer to the up quark, down quark, \ldots\ fields.
The label L refers to left handed Weyl spinors and the upper c symbol 
indicates a charge--conjugated field. The number in parenthesis stands for the
hypercharge and the symbol [$\alpha\beta$] indicates antisymmetric ordering,
$\phi_{[AB]} = \case{1}{\sqrt2}(\phi_{AB}-\phi_{BA})$.
\begin{equation}
\psi_{[\alpha\beta] {\rm L}}(0)=
  \left(
   \begin{array}{cccccc}
    0 &\rm N_1    &\rm E_1^- &\rm N_4   &\rm E_2^- &\rm N_6   \\[1mm]
      &0          &\rm N_5   &\rm E_1^+ &\rm N_7   &\rm E_2^+ \\[1mm]
      &           &0         &\rm N_2   &\rm E_3^- &\rm N_8   \\[1mm]
      &           &          &0         &\rm N_9   &\rm E_3^+ \\[1mm]
      &           &          &          &0         &\rm N_3   \\[1mm]
      &           &          &          &          &0
   \end{array} 
  \right)_{\rm L}                  
\end{equation}
is the  multiplet of exotic leptons. They are classified 
according to $\rm SU(2)_L$ into 3 triplets,
\begin{equation}
 \left( 
  \begin{array}{c} 
   \rm E_1^{+}\\[2mm] \rm \frac{1}{\sqrt2}(N_4+N_5)\\[2mm] \rm E_1^{-}
  \end{array} 
 \right) , \hspace{12mm}
 \left( 
  \begin{array}{c} 
   \rm E_2^{+}\\[2mm] \rm \frac{1}{\sqrt2}(N_6+N_7)\\[2mm] \rm E_2^{-}
  \end{array} 
 \right) , \hspace{12mm}
 \left( 
  \begin{array}{c} 
   \rm E_3^{+}\\[2mm] \rm \frac{1}{\sqrt2}(N_8+N_9)\\[2mm] \rm E_3^{-}   
  \end{array} 
 \right) ,
\end{equation}
and six neutral singlets,
\begin{equation}
 \rm N_1,~~ N_2,~~ N_3,~~ \case{1}{\sqrt2}(N_4-N_5),~~ 
 \case{1}{\sqrt2}(N_6-N_7),~~\case{1}{\sqrt2}(N_8-N_9)~.
\end{equation}
  Notice, from the structure of the 
$\{ {\bf 6}\left( \frac{1}{3} \right) \}_{\rm L}$, that the 
$\rm SU(6)_L$ indices are arranged according to the following scheme:
\begin{equation}
\begin{array}{ c c c c}
               &     \leftarrow    & {\rm SU(3)_H} & \rightarrow  \\[1ex]
 \uparrow      &         1         &       3       &      5       \\
 {\rm SU(2)_L} &                   &               &              \\
 \downarrow    &         2         &       4       &      6       \\
\end{array}
\end{equation}

\subsection{Symmetry Breaking}
\label{symm-break}
The symmetry breaking is realized in three stages: at the scale $M_1$ 
\begin{equation}
 \mbox{$\rm SU(6)_L \!\otimes\! U(1)_Y$}\;\;\stackrel{\!\!M_1}{\longrightarrow}
 \;\;\mbox{$\rm SU(2)_{L} \!\otimes\! SU(2)_{H} \!\otimes\! U(1)_{Y}$}
\end{equation}
and the six $\rm SU(2)_L$ exotic singlets get mass of order $M_1$. 
This breaking is achieved ~\cite{zpc55:423} with a Higgs scalar in the 
irrep $\phi_{1}=\{{\bf105}(0)\}$ and the
horizontal symmetry is reduced to the special $\rm SU(2)_H$ subgroup  
generated by the set 
\begin{equation}
{\sf T}_{\rm H1} =  \case{1}{\sqrt2}({\lambda}_1 + {\lambda}_6) 
  \longleftrightarrow \Sigma_1 , \hspace{8mm} 
{\sf T}_{\rm H2} = \case{1}{\sqrt2}({\lambda}_2 + {\lambda}_7) 
  \longleftrightarrow \Sigma_2 , \hspace{8mm} 
{\sf T}_{\rm H3} = \case{1}{2}({\lambda}_6 + \sqrt{3}{\lambda}_8) 
  \longleftrightarrow  \Sigma_3,
\end{equation}
where the corresponding gauge fields have been indicated.
Here the generators are conveniently normalized to 
Tr${\sf T}_{\rm Hi}{\sf T}_{\rm Hj} = 2$ so 
that ${\sf T}_{\rm H3}$ has eigenvalues $=0, \pm 1$. It then follows that the 
coupling constants of $\rm SU(2)_H$ and $\rm SU(3)_H$ are related, at the 
unification scale, by
\begin{equation}
 g_{\rm 2H} = \case{1}{2} g_{\rm 3H} =  \case{1}{2\sqrt{2}}g_6. \label{g2H}
\end{equation} 
In the second stage of the symmetry breaking chain
\begin{equation}
  {\rm SU(2)_L \!\otimes\! SU(2)_H \!\otimes\! U(1)_Y} 
  \stackrel{\!\!M_2}{\longrightarrow} 
  {\rm SU(2)_L \!\otimes\! U(1)_Y},
\end{equation}
at the scale $M_2$, the exotic leptons which transform as triplets 
of $\rm SU(2)_L$ get a mass of order $M_2$, and ${{\rm E}_i^+}^{\rm c}$ 
becomes the right--handed counterpart of E$_i^-$ . 
 This step is implemented ~\cite{zpc55:423} with a Higgs 
$\phi_{2}=\{\overline{\bf15}(0)\}$. 

 The final stage of the symmetry breaking chain,
\begin{equation}
 {\rm SU(2)_L \!\otimes\! U(1)_Y}
  \stackrel{\!\! M_{\rm Z}}\longrightarrow {\rm U(1)_{EM}},
\end{equation}
is achieved using a Higgs field 
$\phi_{3} = {\overline{\bf 6}(1)}$ with VEV's in the neutral components 
$\langle \phi_{3i} \rangle = v_i$ for i=1,3,5.
 With $\phi_3$ the following mass term for quarks may be written

\begin{equation}                      
 \sum_{I=1}^3\gamma_{I}q^{\rm c}_{I {\rm L}}({-\case{4}{3}}){\rm C 
    \phi_{3\alpha} \psi^{\alpha}_{\rm L}(\case{1}{3}) + h.c.}
     \;\;=\;\;{(\rm \gamma_u u^c + \gamma_c c^c + \gamma_t t^c)_L C} 
     (v_1 {\rm u} + v_2 {\rm c} + v_ 3 \rm t)_L + h.c.\;, 
\end{equation}
where $\gamma_{I}$ are Yukawa couplings of order 1 and $c$ is the charge
conjugation matrix. 
This mass term predicts tree--level zero mass for all the quarks, except for 
the top:
$m_{\rm t}=\gamma v$ where $\rm \gamma=\sqrt{\gamma_u^2+\gamma_c^2+\gamma_t^2}$ 
and $v=\sqrt{v_{1}^{2}+v_{2}^{2}+v_{3}^{2}}$.
A similar mass term for the $\tau$ is avoided postulating a Z$_5$ discrete 
symmetry that distinguishes quarks from leptons. In this way the Yukawa
coupling of $\phi_3$ in the leptonic sector is completely absent
~\cite{zpc55:423}.

\subsection{Mixing of $\protect\bbox{\rm Z^o}$ with heavy bosons} 
\label{Zomix}
Among the 12 neutral gauge bosons related to FCNC, three are relatively light 
(those of $\rm SU(2)_H$ ), with mass of the order of $M_2$. 
The other 9 are heavy, with mass of the order of $M_1$. 
If $\rm Z^o$ mixes with this type of gauge bosons, the mixing with the first 
three will dominate, unless it vanishes as a consequence of some symmetry.

 We are therefore interested in constructing the 
$\rm Z^o$, $\Sigma_1$, $\Sigma_2$, $\Sigma_3$ mass matrix. 
The VEV's of $\phi_2$, 
\begin{equation}
 \langle\phi_2^{[3456]}\rangle = \langle\phi_2^{[1245]}\rangle = - 
 \langle\phi_2^{[1236]}\rangle = V ,
\end{equation}
produce the following $\Sigma_1$--$\Sigma_2$, $\Sigma_3$ mass terms:
\begin{equation}
 {\cal L}_{\Sigma^\pm \Sigma_3}^{\rm m} \simeq
   \case{3}{2}g_6^2 V^2 
 \left( 
  \Sigma^+ \Sigma^- + \case{1}{2} \Sigma_3^2 
 \right), 
\end{equation}
where we do not consider the mixings with the heaviest neutral gauge 
bosons.
$\phi_3$ gives
\begin{equation}
 {\cal L \/}_{\Sigma^\pm \Sigma_3 {\rm Z^o} }^{\rm m} 
 \simeq
  \case{1}{2} \Delta M_{\Sigma_3}^2 \Sigma_3^2
  +\case{1}{2} M_{\rm Z^o}^2 {\rm Z^o}^2
  +\left( 
    \case{1}{2}        M_{\rm Z^o \Sigma^+}^2 {\rm Z^o \Sigma^+} + 
    \case{1}{2} \Delta M_{\Sigma^+}^2 {\Sigma^+}^{2} + {\rm h.c.} 
   \right),
\end{equation}
where
\begin{eqnarray}
 {\rm Z^o}
 &\equiv&
 \frac{g {\rm W^3}-g'{\rm B}}{\sqrt{g^2+g^{\prime 2}}},
\end{eqnarray}
$g$ and $g'$ are the coupling constants of $\rm SU(2)_L$ and $\rm U(1)_Y$ 
respectively,
\begin{equation}
 \rm \Sigma^{\pm} = \case{1}{\sqrt2} 
  \left( 
   \Sigma_1 \mp i\Sigma_2 
  \right),
\end{equation}
and
\begin{equation}
 \Delta M_{\Sigma_3}^2 = \case{1}{4} v^2 g^2,  \hspace{10mm}
 M_{\rm Z^o}^2  = \case{1}{4} v^2(g^2 + g'^2), \hspace{10mm}
 M_{\rm Z^o \Sigma^+}^2 = 
  \case{4}{\sqrt3} v^2 g\sqrt{g^2 + g'^2},     \hspace{10mm} 
 \Delta M_{\Sigma^+}^2 = \case{3}{8} v^2 g^2.
\end{equation}
 Notice that $\Sigma_2$ decouples from $\rm Z^o$ due to the h.c. term and 
that $\Sigma_3$ does not mix with any other. Thus only the 
$\rm Z^o$--$\Sigma_1$ sector must be diagonalized and the mass matrix 
leads to 
\begin{equation}
 \left(
  \begin{array}{c}
   {\rm Z^o}\\
   \Sigma_1
  \end{array}
 \right)
 \;=\;
 \left(
  \begin{array}{rr}
   \cos{\Theta} & -\sin{\Theta}\\
   \sin{\Theta} &  \cos{\Theta}
  \end{array}
 \right)
 \left(
  \begin{array}{l}
   {\rm Z}\\
   {\rm Z}'
  \end{array}
 \right)
 \;\equiv\;
 {\sf R}
 \left(
  \begin{array}{l}
   {\rm Z}\\
   {\rm Z}'
  \end{array}
 \right) ,
\end{equation}
where
\begin{equation}
 \cos{\Theta} \simeq \frac{1}{K_3}, \hspace{5mm}
 \sin{\Theta} \simeq \frac{1}{K_3} \delta
\end{equation}
and
\begin{equation}
 K_3 = \sqrt{1 + \delta^2} , \hspace{10mm}
 \delta = 2 \sqrt{2} \left(  \frac{M_{\rm Z}}{M_{Z'}} \right)^2.
\end{equation}
The mass of the Z gauge boson is given by 
\begin{equation}
 M^2_{\rm Z} \simeq \frac{M^2_{\rm W}}{\cos^2 \theta_{\rm w}} \left[1 -
 \left( \frac{v}{V} \right)^2 \right], 
\end{equation}
where $M_{\rm W}$ is the mass of the charged boson W$^\pm$ and
${\theta}_{\rm w}$ is the weak mixing angle given by
\begin{equation}
 \begin{array}{ccccc}
  g \sin{\theta_{\rm w}} &=& g' \cos{\theta_{\rm w}} &=& e .
 \end{array}
\end{equation}
The mass for the Z$'$ is given by 
\begin{eqnarray}
 M^2_{\rm Z'} &\simeq& \frac{9 g^2}{2} V^2 \; .
\end{eqnarray}

\subsection{Mixing of ordinary charged leptons with exotic ones and family 
            changing couplings}
\label{mixord}
 Let 
$\bbox{\psi}_a^{\rm o} = \left( 
                          \rm e, \mu, \tau, E_1, E_2, E_3,
                         \right)_a^{\rm o \top}$, 
$a = {\rm L, R}$, be a vector of gauge eigenstates in the electric charge
$ q = - 1$ space and $\bbox{\psi}_a$ the one corresponding to the mass 
eigenstates.
 Let 
\begin{equation}
 \bbox{\psi}_a^{\rm o} = {\sf U}_a \bbox{\psi}_a,
\label{fermass}
\end{equation}
with
\begin{equation}
 {\sf U}_a = 
  \left( 
   \begin{array}{cc}
    {\sf A}_a & {\sf E}_a \\[2mm]
    {\sf F}_a & {\sf G}_a  
   \end{array}  
  \right).
\end{equation}
 Let ${\sf D}_a$ and ${\sf H}_a$ be the $6 \times 6$ matrices that express 
the couplings of the $\rm Z^o$ and $\Sigma_1$ to $\bbox{\psi}_a^{\rm o}$,
\begin{eqnarray}
 -\cal{L}^{\rm nc} & = & 
  \frac{e}{\rm s_{\theta_w} c_{\theta_w}}
  \sum_{a={\rm L,R}} \bar{\bbox{\psi}}_a^{\rm o} \gamma^{\mu}  
 \left(
  {\sf D}_a,\; {\sf H}_a
 \right)
 \bbox{\psi}_a^{\rm o}
 \left(
  \begin{array}{l}
   {\rm Z^o}   \\
   {\Sigma_1}  \\
  \end{array}
 \right)_{\mu} ,
\end{eqnarray}
where
\begin{mathletters}
\begin{eqnarray}
{\sf D}_{\rm L}  &=&
  \left(
   \begin{array}{cc}
    -\case{1}{2} + \sin^2 \theta_{\rm w} &     {\sf 0}     \\
             {\sf 0}                     & - 1 + \sin^2 \theta_{\rm w}
   \end{array}
  \right) \otimes {\sf 1}_3 , \\[3mm]
{\sf D}_{\rm R}  &=&
  \left(
   \begin{array}{cc}
    \sin^2 \theta_{\rm w}  &     {\sf 0}                 \\
          {\sf 0}          & - 1 + \sin^2 \theta_{\rm w}
   \end{array}
  \right) \otimes {\sf 1}_3  ,
\end{eqnarray}
\end{mathletters}
\begin{equation}
  {\sf H}_{\rm L} =
  \left(
   \begin{array}{cr}
    \sf h & \sf  0 \\
    \sf 0 & \sf -h 
   \end{array}
  \right), \hspace{10mm}
  {\sf H}_{\rm R} =
  \left(
   \begin{array}{cc}
    \sf 0 & \sf  0 \\
    \sf 0 & \sf  h 
   \end{array}
  \right), \hspace{10mm}
  {\sf h} =
   -\case{\sqrt{3}}{4} g 
  \left(
   \begin{array}{ccc}
    0 & 1 & 0 \\
    1 & 0 & 1 \\
    0 & 1 & 0
   \end{array}
  \right),
\label{HO6}
\end{equation}
then the coupling of the lowest mass neutral gauge boson mass eigenstate,
Z, to the light (ordinary) leptons, $\bbox{\psi}_{la} = (\rm e, \mu,
\tau)_{la}^\top$, is given by \cite{prd55:2998} 
\begin{eqnarray}
 -\cal{L}^{\rm nc}_{\rm Z} & = &
   \frac{e}{\rm \sin_{\theta_w} \cos_{\theta_w}}
    \sum_{a={\rm L,R}} \bar{\bbox{\psi}}_{l a} \gamma^{\mu}
    {\sf K}_a
    \bbox{\psi}_{la}
      {\rm Z}_{\mu} ,
\end{eqnarray}
 where
\begin{mathletters}
\label{coeff} 
\begin{eqnarray}
 {\sf K}_{\rm L} & = &
 \left[
  -\case{1}{2}
  \left( {\sf F^{\dag} F} \right)_{\rm L}
   - \left(
    \case{1}{2} - \sin^2 \theta_{\rm w} \right)
 \right]
 \cos{\Theta} + 
 \left( {\sf H}_{ll} \right)_{\rm L}\sin{\Theta}, \label{KL}\\[3mm]
 {\sf K}_{\rm R} & = & 
 \left[
  -\left(
   {\sf F^{\dag} F} \right)_{\rm R} + \sin^2 \theta_{\rm w}  
 \right]
 \cos{\Theta} +
 \left( {\sf H}_{ll} \right)_{\rm R}\sin{\Theta}, \label{KR}
\end{eqnarray}
\end{mathletters}
\begin{mathletters}
\begin{eqnarray}
 \left(
  {\sf H}_{ll}
 \right)_{\rm L} &=&
 \sf A^\dagger h A - F^\dagger h F, \\[3mm]
 \left(
  {\sf H}_{ll}
 \right)_{\rm R} &=&
 \sf F^\dagger h F. 
\end{eqnarray}
 ${\sf K}_{\rm L}$ and ${\sf K}_{\rm R}$ in eqs.~(\ref{coeff}) may also
by written as 
\end{mathletters}
\begin{mathletters}
\label{coeffdiag}
\begin{eqnarray}
 {\sf K}_{\rm L} & = &
 \left(
  {\Lambda}_{\rm L} - \case{1}{2} + {\rm \sin_{\theta_w}^2}
 \right) \cos{\Theta} +
 {\Xi}_{\rm L} \sin{\Theta}, \label{KeeL}\\[3mm]
 {\sf K}_{\rm R} & = & 
 \left( 
  {\Lambda}_{\rm R} + {\rm \sin_{\theta_w}^2} 
 \right) \cos{\Theta} +
 {\Xi}_{\rm R} \sin{\Theta}, \label{KeeR}
\end{eqnarray}
\end{mathletters}
with
\begin{mathletters}
\begin{eqnarray}
 {\Lambda}_{\rm L} &=& 
  -\case{1}{2}
  \left( 
   {\sf F^{\dag} F} 
  \right)_{\rm L}, \\[3mm]
 {\Lambda}_{\rm R} &=& 
  -\left( 
   {\sf F^{\dag} F} 
  \right)_{\rm R}, \\[3mm]
 {\Xi}_a &=& 
 \left( 
  {\sf H}_{ll} 
 \right)_a . 
\end{eqnarray}
 Another convenient notation is 
${\sf g}_{\rm V} = {\sf K}_{\rm L} + {\sf K}_{\rm R}$, 
${\sf g}_{\rm A} = {\sf K}_{\rm L} - {\sf K}_{\rm R}$.
\end{mathletters}

\subsection{Radiative Charged Lepton Masses}
\label{rad}
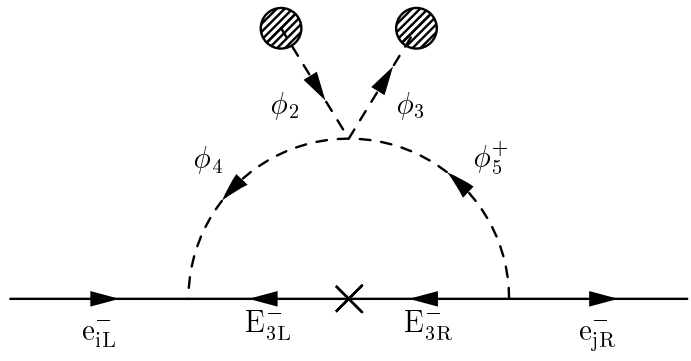
\begin{figure}
 \caption{Diagram generating the $\tau$ mass term}
 \label{mass}
\end{figure}

 In this section we describe a mechanism to generate radiatively the 
masses for the ordinary charged leptons. 
 With the additional Higgs scalars $\phi_4 = \{70(1)\}$ and $\phi_5 =
\{15(-2)\}$ introduced in
Ref.~\cite{zpc55:423}, 
and using as a seed the mass of the exotic charged lepton $E^-_3$, we have 
a contribution coming from the diagrams of Fig.~\ref{mass}, 
where the couplings are given by
\begin{eqnarray}
 \eta \psi^{\top \rm o}_{[\alpha\beta] {\rm L}}(0){\rm C}\psi^{\rm
o}_{\gamma {\rm
L}}(-1)
 \phi^{\alpha\beta\gamma}_4(1) + {\rm h.c.}
 &&
 \\
 + \psi^{\top \rm o}_{[\alpha\beta] {\rm L}}(0) {\rm C}
 \left[
  \sum_I\eta_I l^{c \rm o}_{I {\rm L}}(2)
  \phi^{[\alpha\beta]}_{5 }(-2)
 \right] + {\rm h.c.}
 &&
 \\ 
 + \lambda \phi^+_5 \phi^+_4 \phi_2 \phi^+_3 + {\rm h.c.},
 &&
\end{eqnarray}
where $\eta$ and $\eta_I$ are the Yukawa couplings of order 1.
The mixing of the double charged scalar fields needed in the diagrams
of Fig.1 are generated when $\phi_2$ and $\phi_{3}$ take VEV's.
 The $\Sigma_{ij}(0)$ mass term generated from this diagram is
\begin{equation}
 \Sigma_{ij}(0) 
 = 
 \frac{\lambda_i \lambda_j}{16 \pi^2}{M}_2 
 \sum_l O_{il} O_{jl} f({M}_2, {M}_l)
\end{equation}
where
$\lambda_i$ and $\lambda_j$ are the couplings in the vertices, $O$ is the
unitary matrix which diagonalizes the mass matrix of the scalars,
${M}_l$ 
are the eigenvalues
and
\begin{equation}
 f({M}_2, {M}_l) = \frac{1}{{M}_l^2 - {M}_2^2} 
 \left[
  {M}_l^2  \ln \frac{{M}_l^2}{{M}_2^2} - {M}_l^2 +
  {M}_2^2
   \right]
\end{equation}
 This one loop contribution may be written as $h_i v_j = \Sigma_{ij}(0)$,
and therefore it leads to an increase in one unit of the rank of the
charged leptons mass matrix. 
 In this way the mass of the $\tau$ is generated. 
 To generate the masses for $\mu$ and $e$ we introduce two new
color singlet scalars $\chi_1 = \{21(2)\}$ and
$\chi_2 = \bbox{1}(4)$, and a new Higgs
$\phi_6 = \{21(-2)\}$,
which allow the couplings
\begin{eqnarray}
 H^5 \psi_{\alpha \rm L}^{\top \rm o}(-1) {\rm C} \psi_\beta(-1)_{\rm L}
\chi_1^{\{\alpha\beta\}} + h.c. 
 \\ 
 + H_{Ij}^6 (e^+)^{\top \rm o}_{j \rm L}C(e^+)^{\rm o}_{IL}\chi^+_2 +
h.c.\\
 + \lambda_4 \chi_1^+\chi_2\phi_6 + h.c.
\end{eqnarray} 
  These couplings generate the mass terms in the $e-\mu$ sector  coming from
the diagrams of Fig.~\ref{mass2} and from those diagrams starting with
the right handed chirality from the same Fig.~\ref{mass2}. The appropriate
mixing among the $\chi$ scalars is generated when $\phi_{6}$ takes VEV's.
The contribution from these diagrams can be written again in the form
$h_{i}q_{j}$ and therefore the rank of the charged lepton mass
matrix is again increased in one unit. In this way the muon obtain mass.
Finally the mass of the electron is generated through a diagram similar to
those of Fig.~\ref{mass2}, but with the muon in the internal fermion line.

\section{Constraints}
\label{constr}

\subsection{Family diagonal process 
            $\protect\bbox{\rm Z \rightarrow e \bar{e}}$}
\label{Zlld}

 Mixing effects also modify slightly the rate for family diagonal 
processes~\cite{prd55:2998,prd38:886,npb386:239,prd46:3040}. 
 Consider for example $\rm Z \rightarrow e \bar{e}$, whose branching 
ratio is given by
\begin{mathletters}
\begin{eqnarray}
 {\rm B}(\rm Z \rightarrow e \bar{e}) &\simeq&
 \frac{1}{\Gamma_{\rm tot}} \frac{{\rm G_F} M^3_{\rm Z}}{6 \sqrt{2} \pi}
 \left(
  \left| 
   g_{\rm V}^{\rm ee} 
  \right|^2 +  
  \left| 
   g_{\rm A}^{\rm ee} 
  \right|^2
 \right) \\[3mm]
 &=&
 \frac{1}{\Gamma_{\rm tot}} \frac{{\rm G_F} M^3_{\rm Z}}{3 \sqrt{2} \pi}
 \left(
  \left| 
   \Lambda_{\rm L}^{\rm ee} + \Xi_{\rm L}^{\rm ee} \Theta - 
     \case{1}{2} + {\rm \sin_{\theta_w}^2}
  \right|^2 +  
  \left|  
   \Lambda_{\rm R}^{\rm ee} + \Xi_{\rm R}^{\rm ee} \Theta + {\rm
\sin_{\theta_w}^2}   
  \right|^2
 \right)  + {\rm O}(\Theta^2) \label{Bee}.
\end{eqnarray}
\end{mathletters}
 Since the agreement of the SM predictions with the experimental data for 
these processes is better than 0.1~\% 
(the experimental value of $\Gamma({\rm Z} \rightarrow l \bar{l})$ is 
$83.83 \pm 0.27$~\cite{prd54:1} against the theoretical one equal to 
$83.97 \pm 0.07$), the quantities 
$\Lambda_a^{\rm ee} + \Xi_a^{\rm ee} \Theta$ are bounded practically by the 
experimental uncertainty in the data~\cite{prd54:1},
\begin{eqnarray}
 {\rm B}_{\rm e \bar{e}} \equiv
  {\rm B}(\rm Z \rightarrow e \bar{e}) 
  &=& (3.366 \pm 0.008)  \times  10^{-2}. 
\end{eqnarray}
From eq.~(\ref{Bee}) it follows that
\begin{eqnarray}
\label{values1}
 \left| 
  \Lambda_{\rm L}^{\rm ee} + \Xi_{\rm L}^{\rm ee} \Theta -
    \case{1}{2} + {\rm s_{\theta_w}^2}
 \right|^2 +
 \left| 
  \Lambda_{\rm R}^{\rm ee} + \Xi_{\rm R}^{\rm ee} \Theta + {\rm s_{\theta_w}^2}
 \right|^2
 &=& c {\rm B}_{\rm e \bar{e}},
\end{eqnarray}
where 
$c^{-1} = \frac{1}{\Gamma_{\rm tot}} 
\frac{{\rm G_F} M^3_{\rm Z}}{3 \sqrt{2} \pi} = 0.2675 \pm 0.0005$. 
 Therefore we obtain, in a neighborhood of 
$\left| \Lambda_a^{\rm ee} + \Xi_a^{\rm ee} \Theta \right| = 0$ 
and with $\rm \sin^2_{\theta_w} = 0.2237 \pm 0.0010$,
the bounds
\begin{eqnarray}
 \left| 
  \Lambda_a^{\rm ee} + \Xi_a^{\rm ee} \Theta 
 \right| &<& {\rm few} \times 10^{-3} .
 \label{few}
\end{eqnarray}

\subsection{Constraints from 
            $\protect\bbox{\rm Z \rightarrow e \bar{\mu}}$}
\label{zme}
 With the approximation $M_{\rm Z} \gg m_\mu, m_{\rm e}$ 
and taking into account that experimental limits for 
$\rm Z \rightarrow e \bar{\mu}$, Fig.~\ref{figZme},
exist only for the sum of the charge states of particles and 
antiparticles states, the branching ratio is
\begin{mathletters}
\begin{eqnarray}
 {\rm B}({\rm Z} \rightarrow e \bar{\mu} + \mu \bar{e}) &\simeq&
 2\frac{{\rm B}({\rm Z} \rightarrow l \bar{l})}
 {\left|
   g_{\rm V}
  \right|^2 + 
  \left|
   g_{\rm A}
  \right|^2}
 \left( 
  \left| 
   g_{\rm V}^{\rm e \mu} 
  \right|^2 + 
  \left| 
   g_{\rm A}^{\rm e \mu} 
  \right|^2 
 \right) \\[3mm]
&\simeq&
  4\frac{{\rm B}({\rm Z} \rightarrow 
                  l \bar{l})}{\left|
                               g_{\rm V}
                              \right|^2 + 
                              \left|
                               g_{\rm A}
                              \right|^2} 
 \left(
  \left|
   \Lambda_{\rm L}^{\rm e \mu} + \Xi_{\rm L}^{\rm e \mu} \Theta
  \right|^2 +
  \left|
   \Lambda_{\rm R}^{\rm e \mu} + \Xi_{\rm R}^{\rm e \mu} \Theta
  \right|^2
 \right) + {\rm O}\left( \Theta^2 \right)..
\end{eqnarray}
which leads to 
\end{mathletters}
\begin{eqnarray}
\label{bounds1}
 \left| 
  \Lambda_{\rm L}^{\rm e \mu} + \Xi_{\rm L}^{\rm e \mu} \Theta 
 \right|^2 +
 \left| 
  \Lambda_{\rm R}^{\rm e \mu} + \Xi_{\rm R}^{\rm e \mu} \Theta 
 \right|^2
 &<& c {\rm \widetilde{\rm B}}_{\rm e \bar{\mu}},
\end{eqnarray}
where $c^{-1} = 4\frac{{\rm B}({\rm Z} \rightarrow l \bar{l})}
  {\left| g_{\rm V}\right|^2 + \left|g_{\rm A}\right|^2} = 0.536$
(using the conventional SM branching ratio 0.0337 for B$_{l\bar{l}}$
and the standard values for $g_{\rm V}$ and $g_{\rm A}$) 
and where $\widetilde{\rm B}_{\rm e \bar{\mu}}$ is defined by the numerical 
value of the experimental upper bound~\cite{zpc67:555}
\begin{eqnarray}
 {\rm B}_{\rm e \bar{\mu}} \equiv
   {\rm B}(\rm Z \rightarrow e \bar{\mu} + \mu \bar{e}) 
   &<& 1.7 \times 10^{-6} \equiv 
    \rm \widetilde{B}_{e \bar{\mu}} .
\end{eqnarray}
 This means that the ordinary-ordinary off diagonal mixing parameters 
$\Lambda_a^{\rm e \mu}$ are bounded
to lie in a circular region centered at
$\left( -\Xi_{\rm L}^{\rm e \mu} \Theta, 
-\Xi_{\rm R}^{\rm e \mu} \Theta \right)$  
and of radius $9.5 \times 10^{-4}$.

\subsection{Constraints from $\protect\bbox{\rm \mu \rightarrow ee\bar{e}}$}
\label{meee}
Since $m_{\mu} \gg m_{e}$ and ignoring possible contributions from 
scalars to the process, the branching ratio 
$\rm B(\mu \rightarrow e e \bar{e})$, described by 
Fig.~\ref{figmeee}, is given by
\begin{mathletters}
\begin{eqnarray}
 \frac{\rm B(\mu \rightarrow e e \bar{e})}
      {\rm B (\mu \rightarrow e \bar{\nu}_{e} \nu_{\mu})} &=&
  \frac{1}{2}
  \left[ 3
   \left(
    \left|g^{\rm ee}_{\rm V}\right|^2 + \left|g^{\rm ee}_{\rm A}\right|^2
   \right) 
   \left(
    \left|g^{\rm e \mu}_{\rm V}\right|^2 + \left|g^{\rm e \mu}_{\rm A}\right|^2
   \right) + \right. \nonumber \\[2mm] && \left. \hspace{18mm}
   2\Re e
   \left(
    g^{\rm ee}_{\rm V} {g^{\rm ee}_{\rm A}}^\ast
   \right)   
   2\Re e
   \left(
    g^{\rm e \mu}_{\rm V} {g^{\rm e \mu}_{\rm A}}^\ast
   \right)
  \right] + \nonumber \\ &&
   \frac{M^2_{\rm Z}}{M^2_{\rm Z'}}
  \Re e
  \left[
  3
  \left(
   g^{\rm ee}_{\rm V} {g^{\prime \rm ee}_{\rm V}}^\ast + 
   g^{\rm ee}_{\rm A} {g^{\prime \rm ee}_{\rm A}}^\ast
  \right)
  \left(
   g^{\rm e \mu}_{\rm V} {g^{\prime \rm e \mu}_{\rm V}}^\ast + 
   g^{\rm e \mu}_{\rm A} {g^{\prime \rm e \mu}_{\rm A}}^\ast
  \right) + \right. \nonumber \\[2mm] && \left. \hspace{18mm}
  \left(
   g^{\rm ee}_{\rm V} {g^{\prime \rm ee}_{\rm A}}^\ast + 
   g^{\rm ee}_{\rm A} {g^{\prime \rm ee}_{\rm V}}^\ast
  \right)
  \left(
   g^{\rm ee}_{\rm V} {g^{\prime \rm ee}_{\rm A}}^\ast + 
   g^{\rm ee}_{\rm A} {g^{\prime \rm ee}_{\rm V}}^\ast
  \right)
  \right] + \nonumber \\ &&
   \frac{1}{2} \frac{M^4_{\rm Z}}{M^4_{\rm Z'}}
  \left[ 3
   \left(
    \left|g^{\prime \rm ee}_{\rm V}\right|^2 + \left|g^{\prime \rm ee}_{\rm A}\right|^2
   \right) 
   \left(
    \left|g^{\prime \rm e \mu}_{\rm V}\right|^2 + \left|g^{\prime \rm e \mu}_{\rm A}\right|^2
   \right) + \right. \nonumber \\[2mm] && \left. \hspace{18mm}
   2\Re e
   \left(
    g^{\prime \rm ee}_{\rm V} {g^{\prime \rm ee}_{\rm A}}^\ast
   \right)   
   2\Re e
   \left(
    {g^{\prime \rm e \mu}_{\rm V}} {g^{\prime \rm e \mu}_{\rm A}}^\ast
   \right)
  \right] 
\label{explain} \\[3mm]
&\simeq&
 4\left[
  \left(
   2\left|
    -\case{1}{2} + {\rm s_{\theta_w}^2}
   \right|^2 +
   \left|
`    {\rm s_{\theta_w}^2}
   \right|^2 
  \right)
   \left|
    \Lambda^{\rm e \mu}_{\rm L} + \Xi^{\rm e \mu}_{\rm L} \Theta 
   \right|^2 + \right. \nonumber \\[2mm] && \left. \hspace{18mm}
  \left(
   \left|
    -\case{1}{2} + {\rm s_{\theta_w}^2}
   \right|^2 +
   2\left|
    {\rm s_{\theta_w}^2}
   \right|^2 
  \right)
   \left|
    \Lambda^{\rm e \mu}_{\rm R} + \Xi^{\rm e \mu}_{\rm R} \Theta 
   \right|^2   
 \right]  + {\rm O}\left( \Theta^2\right),
\end{eqnarray} 
 where we have assumed
$\left(\frac{M_{\rm Z}}{M_{\rm Z'}}\right)^2 \sim \Theta$,
$\Lambda_a^{\rm e \mu} \lesssim \Theta$
(remember that 
$\Lambda_a^{\rm e \mu}$ is second order in the ordinary-exotic mixing)
and we have taken into account the stringent
limits obtained in eq.~(\ref{few}) from which 
\end{mathletters}
\begin{mathletters}
\begin{eqnarray}
 \left|
  \Lambda_{\rm L}^{\rm ee} + \Xi^{\rm ee}_{\rm L} \Theta
    -\case{1}{2} + {\rm s_{\theta_w}^2}
 \right| &\simeq&
 \left|
  -\case{1}{2} + {\rm s_{\theta_w}^2}
 \right|, \\[3mm]
 \left|
  \Lambda_{\rm R}^{\rm ee} + \Xi^{\rm ee}_{\rm R} \Theta + {\rm s_{\theta_w}^2}
 \right| &\simeq&
 \left|
  {\rm s_{\theta_w}^2}
 \right|.
\end{eqnarray}
 Using the experimental limits~\cite{npb299:1}    
\end{mathletters}
\begin{eqnarray}
 \rm B_{\mu e e \bar{e}} \equiv
     B(\mu \rightarrow e e \bar{e})
 &<& 1.0 \times 10^{-12} \equiv
    \rm \widetilde{B}_{\mu e e \bar{e}}     
\end{eqnarray}
and 
$\rm s^2_{\theta_w} = 0.2237$, 
the constraints on the mixing parameters are
\begin{eqnarray}
 \label{bounds2} 
 0.203 \left| \Lambda_{\rm L}^{\rm e \mu} + \Xi_{\rm L}^{\rm e \mu} \Theta \right|^2 +
 0.176 \left| \Lambda_{\rm R}^{\rm e \mu} + \Xi_{\rm R}^{\rm e \mu} \Theta \right|^2
 &<& c_{\mu} {\rm \widetilde{B}}_{\rm \mu e e \bar{e}}
\end{eqnarray}
where
$c_{\mu} = \left(
4 \rm B \left( \mu \rightarrow e \bar{\nu}_{e} \nu_{\mu}\right) \right)^{-1} 
 \approx 0.25$.
Eq.~(\ref{bounds2}) is more stringent than eq.~(\ref{bounds1}).

 A possible contribution from scalars to this process come from the
fermionic mass-generation mechanism through a complete penguin diagram as
in the Fig.~\ref{fig1meg}.

\subsection{Constraints from $\protect\bbox{\rm \mu \rightarrow
    e\gamma}$}
\label{meg}
 In this section we analyze the lepton flavor violation process 
$\rm \mu \rightarrow e\gamma$ arising in the model mainly from the 
$Z^0 - \Sigma_1$ mixing and from the effect induced of the fermionic 
mass-generation mechanism at one loop level. 
 
 The contribution to this process coming from the mixing of ordinary with
exotic leptons is negligible compare with the above mention sources. This
is a consequence of the approximations obtained in eqs.(54a) and (54b),
because these results means that the modifications to the couplings of the
Z and Z$'$ to the charged leptons are not sensitive to the mixing of
ordinary with exotic leptons. In the other hand when we write the gauge
eigenstates in terms of the mass eigenstates, eq.(28), the dominant
contribution come from the diagonal elements of the $U$ matrix, that is we
are working in the limit of very small off-diagonal elements of the
matrix $U$.

 Contributions from the would be Goldstone bosons are forbidden in the 
model.
 The reason is that the amplitude for 
$\rm \mu \rightarrow e\gamma$ 
is of the form 
$\bar{u}_2 ({p_2}) \sigma^{\mu \nu} q_{\nu}
\epsilon_{\mu} \bar{u}_1 ({p_1})$, 
where 
${q} = {p_1 - p_2}$ 
and 
$\epsilon_\mu$ 
is the polarization of the photon.
 The process involves therefore a flip of helicity
and the lowest order contribution would arise from  diagrams of the
type of Fig.~\ref{fig2meg}.
 However, as was mentioned in Section~\ref{symm-break},
$\phi_3$ is forbidden, by a $Z_{5}$ discrete symmetry, to couple to
leptons.

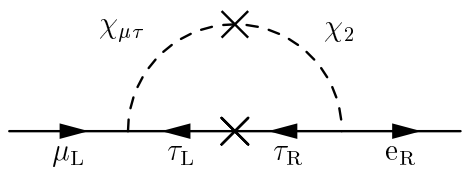
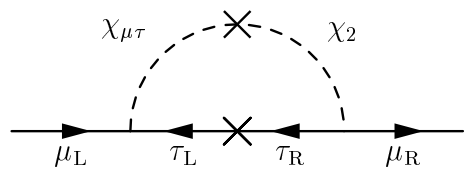
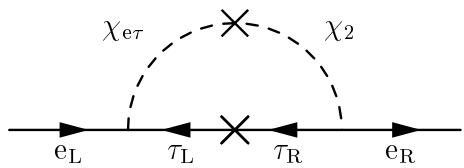
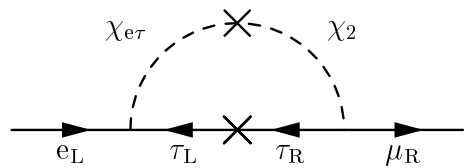
\begin{figure}
 \caption{Diagrams generating mass terms in the $e - \mu$ sector}
 \label{mass2}
\end{figure}

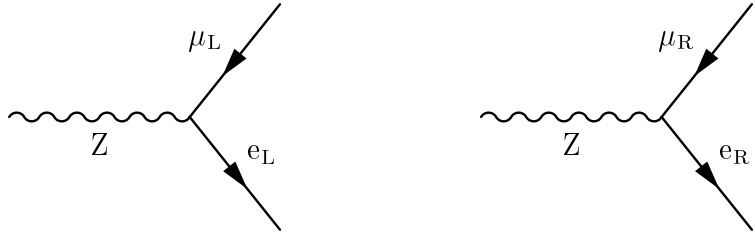
\begin{figure}
 \caption{Diagrams contributing to the $\rm Z \rightarrow e \bar{\mu}$ decay.}
 \label{figZme}
\end{figure}

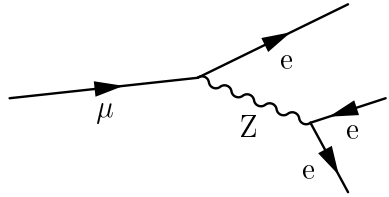
\begin{figure}
 \caption{Contribution to the $\rm \mu \rightarrow ee\bar{e}$ process.}
 \label{figmeee}
\end{figure}

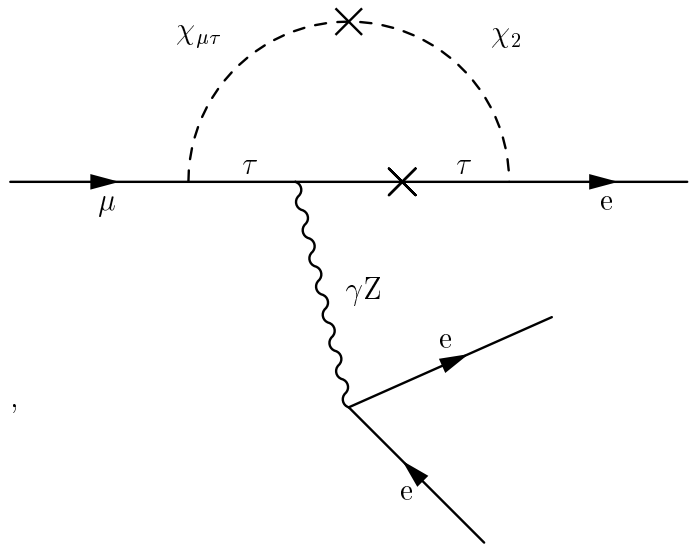
\begin{figure}
 \caption{Induce contribution to the  
          $\rm \mu \rightarrow ee\bar{e}$ process coming from the mass
generation mechanism.}
 \label{fig1meg}
\end{figure}

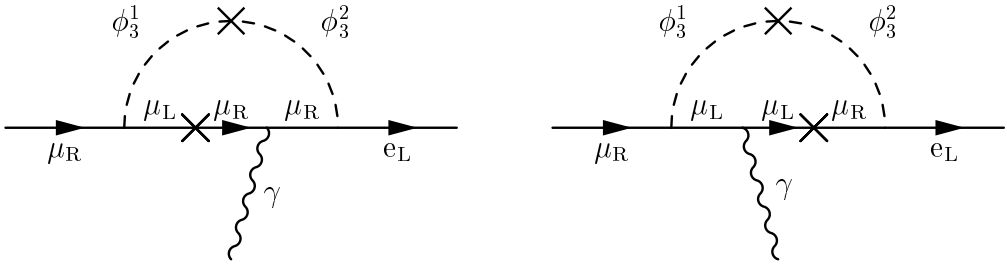
\begin{figure}
 \caption{Diagrams forbidden by the Z$_5$ discrete symmetry.}
 \label{fig2meg}
\end{figure}

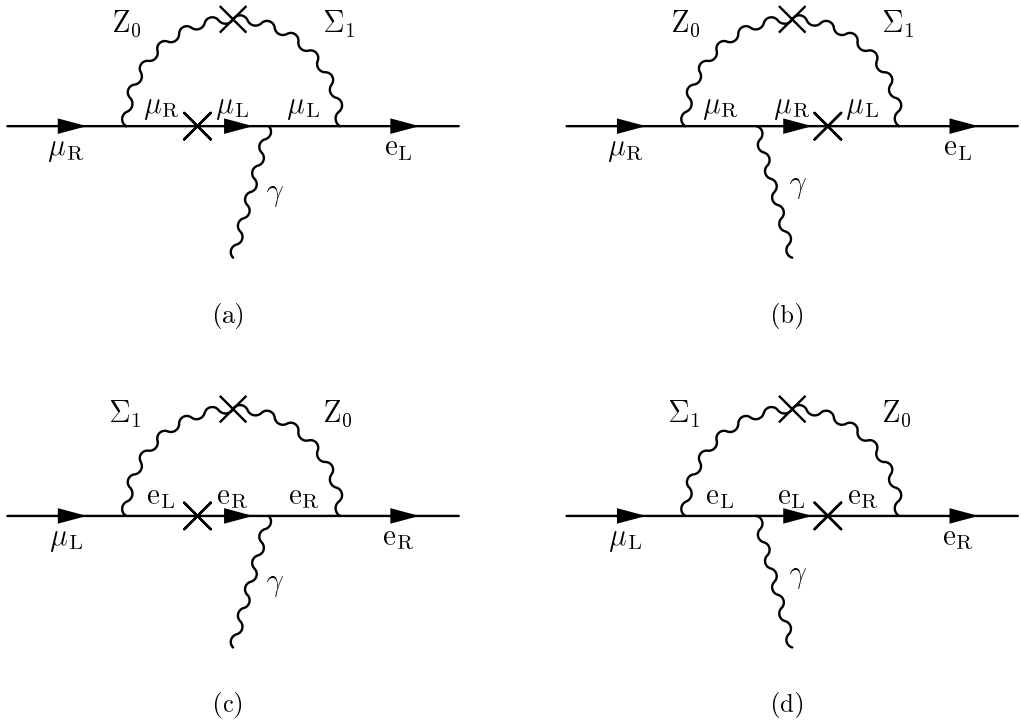
\begin {figure}
 \caption{contribution to the $\rm \mu \rightarrow e \gamma$ decay coming
from the $Z^{0} - \Sigma_{1}$ mixing.}
 \label{fig3meg}
\end{figure}

 Contributions from $Z^0 - \Sigma_1$ mixing are given through the
diagrams of Fig.~\ref{fig3meg}, and from the mechanism of mass generation
by inserting
a photon in the internal lines of the diagram of Fig.~\ref{mass2}a, and
the
conjugate diagram of Fig.~\ref{mass2}d, which produce off-diagonal mass
terms in
the $e-\mu$ sector. 
 Note that contributions from the diagrams of Fig.~\ref{fig3meg}a
and Fig.~\ref{fig3meg}b are proportional to the mass of the muon, while
the
contribution
from diagrams of Fig.~\ref{fig3meg}c and Fig.~\ref{fig3meg}d are
proportional to the mass of the
electron. 
 That is, the dominant contribution from the $Z^0 - \Sigma_1$
mixing is to $\rm \mu_{R} \rightarrow e_{L}\gamma$.
 By simplicity and economy we align the VEV´s of $\phi_6$ such that
$\chi_2$ mixes only with $\chi_{\mu\tau}$. In this case, the contribution
from the mass-mechanism is of the form 
$\rm \mu_L \rightarrow e_R \gamma$ through the diagram of
Fig.~\ref{mass2}a. 
 The calculation of the diagram of Fig.~\ref{mass2}a without the photon
yields the
result
\begin{equation}
 \label{eq:aaa}
 m_{12}=\frac{\lambda_1 \lambda_2}{16 \pi^2} m_\tau \ln
\frac{{M}_1^2}{{M}_2^2}
 \cos\alpha \sin\alpha \sim m_\mu,
\end{equation}
where $\lambda_{1}$, $\lambda_{2}$ are the couplings of the vertices,
\begin{equation}
\left(
 \begin{array}{c}
  {\chi_2}\\
  {\chi_{\mu\tau}}
  \end{array}
 \right)
\;=\;
\left(
  \begin{array}{rr}
    \cos\alpha & -\sin\alpha \\
    \sin\alpha &  \cos\alpha
    \end{array}
    \right)
\left(
 \begin{array}{l}
  {\chi_{2}^{\prime}}\\
  {\chi_{\mu\tau}^{\prime}}
   \end{array}
    \right),
\end{equation}

${M}_1$ and ${M}_2$ are the mass eigenvalues,
$\chi_{2}^{\prime}$ and $\chi_{\mu\tau}^{\prime}$ are the mass
eigenstates, with the approximation ${M}_1,{M}_2 \gg m_\tau$.

 The amplitudes obtained by computing the one loop diagrams are
\begin{mathletters}
\begin{eqnarray}
 i {\rm A_{B}} &=&
   \left(
    C_{B} \bar{e} i
    \right)
     \frac{\sigma_{\mu \nu} q_{\nu} \epsilon_{\mu}}{m_{\mu} + m_e}
      \frac{1 + \gamma_5}{2} \mu  
\end{eqnarray}
\end{mathletters}
and
\begin{mathletters}
\begin{eqnarray}
 i {\rm A_{M}} &=&
    \left(
     C_{M} \bar{e} i
     \right)
      \frac{\sigma_{\mu \nu} q_{\nu} \epsilon_{\mu}}{m_{\mu} + m_e}
       \frac{1 - \gamma_5}{2} \mu,
\end{eqnarray}
\end{mathletters}
 
 where ${\rm A_{B}}$ comes from the $Z^0 - \Sigma_1$ mixing, ${\rm A_{M}}$
from the mass mechanism and the coefficients $C_{B}$ and $C_{M}$ are given
by
\begin{mathletters}
\begin{eqnarray}
C_{B} &=&
        \frac{5 e}{12 \sqrt{2} \pi} 
         \alpha m_{\mu}
          \left(m_{\mu} + m_e
           \right)
            \frac{{\sin\Theta} {\cos\Theta}}                   
             {{\cos\theta_{w}} M^2_{Z}}
\end{eqnarray}
and
\begin{eqnarray}
C_{M} &=&
      \frac{e}{{M}_1^2} m_{\mu}
       \left(m_{\mu} + m_e
        \right)       
         \frac{1}{\ln \frac{{M}_1^2}{{M}_2^2}}
          \left[\ln \frac{{M}_1^2}{m_{\tau}^2} - 
           \frac{{M}_1^2}{{M}_2^2} \ln
            \frac{{M}_2^2}{m_{\tau}^2} 
             \right]
\end{eqnarray}
\end{mathletters}

 Comparing with the general expressions
\begin{mathletters}
\begin{eqnarray}
 i{\rm m
   \left[
    f_1(p_1) \rightarrow f_2(p_2) + \gamma(q)
     \right]} &=& 
      \left(
       \bar{u_2}(p_2) i
        \right) 
         \frac{\sigma_{\mu \nu} q_{\nu} \epsilon_{\mu}}{m_1 + m_2}
          \left[
             F(0)^V_{21} + F(0)^A_{21} \gamma_5
             \right]
              u_1(p_1), \\[3mm]
{\rm B(\mu \rightarrow e\gamma)} &=&
  \frac{m_{\mu}}{8 \pi}
   \left(
    1 - \frac{m_e}{m_{\mu}}
    \right)^2
     \left(
      1 - \frac{m_e^2}{m_{\mu}^2}
      \right)
       \left[
        \left|
         F(0)^V_{21} \right|^2 +
         \left|
          F(0)^A_{21} \right|^2
          \right]
\end{eqnarray}
\end{mathletters}
for a process $f_1 \longrightarrow f_2 + \gamma$ for a real photon, from
ref.~\cite{prd16:1444}. In our case 
\begin{mathletters}
\begin{eqnarray}
 F(0)^V_{21} &=&
    \frac{1}{2}
     \left(
      C_{M} + C_{B}
      \right), \\[3mm]
 F(0)^A_{21} &=&
      \frac{1}{2}
       \left(
        C_{B} - C_{M}
        \right)
\end{eqnarray}
\end{mathletters}
and in the limit $m_{\mu} \gg m_e$ we get  
\begin{mathletters}
\begin{eqnarray}
\label{Bmeg}
 {\rm B(\mu \rightarrow e\gamma)} &=&
   \frac{m_{\mu}}{16 \pi}
    \left(
     C_{B}^2 + C_{M}^2
      \right)
\end{eqnarray}
\end{mathletters}    
 In the approximation ${M}_1\sim {M}_2$ and using the
experimental
limit~\cite{prd38:2077}
\begin{equation}
 \rm B_{\mu e \gamma} \equiv B(\mu \rightarrow e \gamma) <
        4.9 \times 10^{-11} \equiv \widetilde{\rm B}_{\mu e \gamma},
\end{equation}
the constraints on the mixing parameters are $|\Theta|< 4.2\times
10^{-5}$, 
$M_{Z'}> 24$ TeV and $M_1> 200$ TeV. 

\section{Conclusions}
 In this article we have analyzed the consequences of simultaneous
mixing of the Z gauge boson with an horizontal neutral gauge boson 
and the mixing of ordinary charged leptons with exotic ones in the
context of the model $\rm SU(6)_L \otimes U(1)_Y$. 
 We concentrated on the effects of the above mixing on lepton family
violation processes in the e--$\mu$ sector. The charged leptons in this
model obtain mass radiatively by introducing new exotic scalar particles.
This new scalars also contribute to lepton violation processes. In order
to be consistent with experiments we find that their masses must be
heavier than $200$ TeV, $|\Theta|< 4.2 \times 10^{-5}$, and $M_{Z'}> 24$
TeV. We also determine that: 
 i) Since $\Xi^{\rm e \mu}_a$ are order O(1) and $\Theta$ is very small, the 
$\Lambda^{\rm e \mu}_a$ are bounded by each one of 
the processes to lie in a circular region centered in 
$(\rm -\Xi_L^{e \mu}, -\Xi_R^{e \mu})$ and with radius depending on the 
considered branching ratio, 
 ii) The radius fixed by eq.~(\ref{bounds2}) is  
$2.2 \times 10^{-6}$.

\acknowledgements
 This work was partially supported by CONACyT in Mexico and
COLCIENCIAS in Colombia. One of us (A.Z.) acknowledges the hospitality
of Prof. J. Bernabeu and of the Theory Group at the University of 
Valencia as well as the financial, support during the 1995-1996 sabbatical 
leave, of Direcci\'on General de Investigaci\'on Cient\'{\i}fica y T\'ecnica
(DGICYT) of the Ministry of Education and Science of Spain as well as the
hospitality and financial support of the Institute of Nuclear Theory in 
Seattle during part of the summer of 1995.
 One of us (U.C.) would like to acknowledge Gabriel L\'opez for valuable 
discussion.



\end{document}